\documentclass[a4,11pt]{cip-submit-2019}  
\usepackage{color,soul}
\usepackage{afterpage}

\usepackage{wrapfig}
\usepackage{wasysym}
\usepackage{enumitem}
\usepackage{adjustbox}
\usepackage{ragged2e}
\usepackage[svgnames,table]{xcolor}
\usepackage{tikz}
\usepackage{longtable}
\usepackage{changepage}
\usepackage{setspace}
\usepackage{hhline}
\usepackage{multicol}
\usepackage{tabto}
\usepackage{float}
\usepackage{multirow}
\usepackage{makecell}
\usepackage[T1]{fontenc}
\usepackage[utf8]{inputenc}
\usepackage{lmodern}

\usepackage{mathptmx}
\DeclareSymbolFont{epsilon}{OML}{cmm}{m}{it}
\DeclareMathSymbol{\epsilon}{\mathord}{epsilon}{"0F}
\def\DD{D\kern-.7em\raise0.25ex\hbox{\char '55}\kern.33em}
\usepackage{hyperref}
\hypersetup{
	colorlinks=false,
	pdfborder={0 0 0}
}
\usepackage{graphicx,subfigure,wrapfig,epstopdf}
\def\Mr{\uppercase}
\usepackage{graphics}
\usepackage{wasysym}
\usepackage{amsmath,amsxtra,amssymb,latexsym,amscd,color,cite,bm}
\usepackage[mathscr]{eucal}
\usepackage{array}
\usepackage{multirow}
\usepackage{cite} 

\def\AA{nucleus-nucleus\ }

\def\doi#1{\href{http://dx.doi.org/#1}{doi:#1}}
\def\titles#1{\title{\large\bf\noindent #1}}
\def\authors#1{\author{\begin{flushleft}{#1}\end{flushleft}}}
\def\authord#1#2{\indent\Mr{#1}$^{#2}$}
\def\addressed#1#2{\\[1mm]\textit{$\!\!\!^{#1}$\indent#2}}

\def\Email{$^{\dagger}$}
\def\email#1{\bigskip\href{mailto:#1}{\!\!\Email\textit{E-mail:}~{#1}}\\[3mm]}

\def\Keywords#1{\\[.2cm] \textnormal{Keywords:~{#1}}.} 
\def\AND{$\text{\Small AND }$}
\def\and{$\text{\tiny AND }$}

\def\Classification#1{\\[.2cm] \textnormal{Classification numbers:~{#1}.}}  

\newcommand{\sout}[1]{\unskip}

\def\ed{
	\bibliographystyle{cip-sty-2019}
	\bibliography{references-database-name}
\end{document}
}
\begin{document}
	\Year{2021}
	\Page{1}\Endpage{9}
	\titles{Ambiguities from nuclear interactions \\ in the $^{12}$C($p,2p$)$^{11}$B reaction}
	\authors{
	\authord{Nguyen Tri Toan Phuc}{1,2} \Email, \authord{Kazuyuki Ogata}{3,4,5}, \authord{Nguyen Hoang Phuc}{6}, \\ \authord{Bui Duy Linh}{6}, \authord{Vo Hong Hai}{1,2}  \AND \space 
	\authord{Le Xuan Chung}{6}
	\newline
	\addressed{1}{Department of Nuclear Physics, Faculty of Physics and Engineering Physics, 
	 \\ University of Science, VNU-HCM, Ho Chi Minh City, Vietnam.}
	\addressed{2}{Vietnam National University, Ho Chi Minh City, Vietnam.}
	\addressed{3}{Research Center for Nuclear Physics (RCNP), Osaka University,
		Ibaraki 567-0047, Japan.}
	\addressed{4}{Department of Physics, Osaka City University, Osaka 558-8585,
		Japan.}
	\addressed{5}{Nambu Yoichiro Institute of Theoretical and Experimental Physics
		(NITEP), Osaka City University, Osaka 558-8585, Japan.}
	\addressed{6}{Institute for Nuclear Science and Technology, VINATOM \\ 
	 179 Hoang Quoc Viet, Cau Giay, Hanoi, Vietnam}	\\	
	\email{nttphuc@hcmus.edu.vn} 
	}
	\maketitle
	\markboth{Ambiguities from nuclear interactions in the $^{12}$C($p,2p$)$^{11}$B reaction}{N.~T.~T. PHUC, K. OGATA, N.~H. PHUC, B.~D. LINH, V.~H. HAI, AND \space L.~X. CHUNG}

\begin{abstract}
We investigate the impact of ambiguities coming from the choice of optical potentials and nucleon-nucleon scattering cross sections on the spectroscopic factors extracted from the $^{12}$C($p,2p$)$^{11}$B reaction. These ambiguities are evaluated by analyzing the cross sections of the $^{12}$C($p,2p$)$^{11}$B reaction at 100 and 200 MeV within the framework of the distorted-wave impulse approximation with realistic choices of nuclear inputs. The results show that the studied ambiguities are considerably large in this energy region and careful choices of nuclear inputs used in the reaction calculations are required to extract reliable structure information. 
\Keywords{DWIA, knockout, spectroscopic factor, optical potential}
\Classification{24.10.Eq, 21.10.Jx, 25.70.Bc}
\end{abstract} 

\section{\Mr{Introduction}}

The proton-induced nucleon knockout reaction of the $(p,pN)$ type is a powerful spectroscopic tool to probe the single-particle properties of the nucleus such as the separation energy, angular momentum, and spectroscopic factor (SF) \cite{Jac66,Jac73,Kit85,Wak17}. Although $(e,e'p)$ reactions can be used to study some of these properties with better precision, with current experimental capabilities they cannot be applied to investigating unstable nuclei or neutron single-particle states \cite{Suda17}. On the other hand, such limitations have already been overcome with $(p,pN)$ reactions \cite{Aum21}. It is well known that under a proper kinematic condition usually with an incident energy $>200$ MeV/nucleon, the extracted quantities from proton-knockout $(p,2p)$ reactions generally agree with those from $(e,e'p)$ reactions \cite{Kit85,Wak17}. Moreover, $(p,pN)$ reactions with polarized beams can also be used to identify the total angular momentum $j$ of the struck nucleons. The proton-induced knockout reactions can be reliably described within the framework of the distorted-wave impulse approximation (DWIA) \cite{Jac66,Jac73,Kit85,Wak17}. This method provides a consistent description of knockout reactions at forward (normal) and inverse kinematics with flexibility in the inclusion of various nuclear inputs and corrections.  

In recent years, thanks to the latest technological improvements, $(p,pN)$ experiments can be performed with unstable beam in inverse kinematics at several rare isotope beam (RIB) facilities \cite{Aum21,Kob08,Obe11,Naka17,Pan21,Pat21,Paul19,Fro20}. These reactions have been applied to studying various problems from the quenching of nucleon SFs \cite{Aum21,Pan21,Kaw18} to the evolution of single-particle structures in a wide variety of exotic nuclei from light \cite{Tang20,Kub20,Yang21} to medium ones \cite{Oli17,Chen19,Bro21,Tan19}. Many of these applications of $(p,pN)$ reactions require an accurate theoretical description in order to extract reliable results \cite{Aum21}. Although the validity of the DWIA model has been firmly confirmed in the past with normal kinematic experiments, its reliability has been known to depend to some extent on the choice of various nuclear inputs used in the reaction calculation, especially the nuclear interactions \cite{Kit85,Wak17}, which comprise the optical potential (OP) and nucleon-nucleon ($NN$) scattering cross section, are the largest sources of ambiguity in almost all direct nuclear reactions. 

With the increasingly demanding requirement of high accurate cross sections in modern $(p,pN)$ analyses \cite{Phu19,Phuc21}, several aspects of the calculation process have been reevaluated in recent years \cite{Yos16,Yos18,Phu19,Mec19,Gom20,Aum21}. However very few $(p,pN)$ studies extensively investigate the impact of nuclear interactions on the extracted SF, especially for reactions with light nuclei (see Ref.~\cite{Wak17}). Due to the limited availability of high-quality global phenomenological OPs and $NN$ scattering cross sections in the past, the ambiguities in $(p,pN)$ reactions associated with these ingredients were often studied using a restricted set of OP and $NN$ interaction. This restricted nature of the considered ingredients may lead to an underestimation of theoretical ambiguities in the final results. Moreover, Ref.~\cite{Phu19} has suggested a nonnegligible existence of high-order processes at large recoil momentum $p_R>150$ MeV/$c$, which may significantly affect the extracted results from $(p,pN)$ reactions at several RIB facilities. An accurate estimation for these high-order processes requires precise knowledge of ambiguities from the nuclear inputs, to which the distorting OP and the $NN$ scattering cross section contribute a large part.                

In this work, we perform an analysis of $^{12}$C($p,2p$)$^{11}$B reaction at 100 and 200 MeV using the partial-wave DWIA method. The initial analyses of the experimental data at these energies \cite{Cow89,Dev79} use less optimal choices of OPs and $NN$ scattering cross sections in terms of variety and quality. By analyzing these data with several consistent choices of these two inputs, we attempt to study the ambiguities of the extracted $p$-state SFs coming from them. For this purpose, we consider several global Schr\"{o}dinger- and Dirac-based phenomenological OPs, as well as microscopic single-folding potentials, and three different sets of $NN$ scattering cross sections. The energy range in the lower limit of the quasifree impulse approximation considered in this study allows direct comparisons of Schr\"{o}dinger-based potentials with Dirac-based one. The differences between those potentials are also expected to be enhanced due to the strong absorption in this energy region. The extracted SFs along with their uncertainties are compared with those from the same $^{12}$C($p,2p$)$^{11}$B reaction measured at RCNP and using the same calculation ingredients.

\section{THEORETICAL FORMALISM}
\label{sec:theoform}
      
\subsection{Distorted-wave impulse approximation}      
\label{sec:DWIA}

In this section, we briefly introduce the partial-wave DWIA framework. Details of the current DWIA formalism are presented in \cite{Wak17,Yos16,Oga15}. We label the incident proton as particle 0, and the outgoing protons as particles 1 and 2. Quantities with the superscript L are evaluated in the laboratory frame while those without the superscript are evaluated in the three-body center-of-mass (c.m.) frame, or the G frame. 

The triple differential cross section (TDX) of the $A(p,2p)B$ reaction in the factorized form of the DWIA framework is

\begin{equation}\label{eq:tdx}
	\frac{d^3 \sigma}{dE_1^{\rm L} d\Omega_1^{\rm L} d\Omega_2^{\rm L}}
	=S F_{\rm kin}C_0
	\frac{d\sigma_{pp}}{d\Omega_{pp}}
	\left| \bar{T}_{{\bm K}_0{\bm K}_1{\bm K}_2} \right| ^2,
\end{equation}
where 
\begin{equation}\label{eq:c0}
	C_0 = \frac{1}{2(2l+1)}
	\frac{E_0^\mathrm{L}}{(\hbar c)^2 K_0^\mathrm{L}}
	\frac{\hbar^4}{(2\pi)^3 \mu_{pp}^2},
\end{equation}
and 
\begin{equation}\label{eq:fkin}
	F_{\rm kin}= J_{G\to L} \dfrac{K_1 K_2 E_1 E_2}{(\hbar c)^4}\left[1+ \dfrac{E_2}{E_B} +\dfrac{E_2}{E_B} \dfrac{\bm{K}_1 \cdot \bm{K}_2}{K^2_2}  \right]^{-1},
\end{equation}
with $S$ being the proton SF of A nucleus, $\mu_{pp}$ the reduced mass of the two-proton system, and $J_{G\to L}$ is the Jacobian for the transformation from the G frame to the L frame. Although the DWIA framework in this study is nonrelativistic, we adopt the relativistic kinematics through the uses of energy-momentum relation $E_i$=$\sqrt{(m_i c^2)^2+(\hbar c K_i)^2}$ and Lorentz transformation between the asymptotic momenta $\bm{K}_i$. 

The reduced transition amplitude, also called the distorted momentum distribution is given by
\begin{equation}\label{eq:transam}
	\bar{T}_{{\bm K}_0{\bm K}_1{\bm K}_2}
	=\int d\bm{R}\,	\chi_{1,{\bm K}_1}^{(-)}({\bm R})
	\chi_{2,{\bm K}_2}^{(-)} ({\bm R})
	\chi_{0,{\bm K}_0}^{(+)}({\bm R})\varphi_{p}(\bm{R})
	e^{-i\bm{K}_0\cdot\bm{R} / A},
\end{equation}
where $\chi_{i,\bm{K}_i}$ are the distorted scattering wave functions of the $p$-$A$ ($i=0$) and $p$-$B$ ($i=1,2$) systems, while the superscripts $(+)$ and $(-)$ characterize the outgoing and the incoming boundary conditions of these scattering waves, respectively. One of the main subjects of our investigation is the distorted OP used in the Schr\"{o}dinger equation to generate these scattering wave functions. The normalized bound state wave function of the struck proton is denoted as $\varphi_{p}$. 

In the $(p,2p)$ reaction, the $pp$ scattering cross sections in Eq.~(\ref{eq:tdx}) are in principle half-off-the-energy-shell. However, we can relate this cross section with the measurable one by using a specific on-the-energy-shell (or on-shell) approximation. The on-shell cross section in the G frame is related to the $t$ matrix for $NN$ scattering via

\begin{equation} \label{eq:NNxsec}
\frac{d\sigma_{pp}}{d\Omega_{pp}}(E_{pp},\theta_{pp})\approx \dfrac{M_{pp}^2}{{(2\pi\hbar^2)}^2} |\langle \bm{\kappa}' |t_{pp}| \bm{\kappa}\rangle|^2,
\end{equation}  
where $M_{pp}^2$ is the reduced energy,  $\bm{\kappa}$ and $\bm{\kappa}'$ are the $pp$ relative momenta in the initial and final channels, respectively. Depending on the type of on-shell approximation, the $pp$ scattering energy can be determined accordingly. Three commonly used on-shell approximations are the final-, initial-, and average energy prescriptions \cite{Red70}. 

The $pp$ scattering cross section $d\sigma_{pp}/d\Omega_{pp}$ in Eqs.~(\ref{eq:tdx}) and (\ref{eq:NNxsec}) is defined in the $p$-$A$ c.m.~ frame (G frame). By default, the $NN$ scattering cross sections are defined in the two-nucleon c.m.~ frame (t frame). The cross section in G frame is related to the one in the t frame through the relativistic M\o ller factor $\eta$ \cite{Mol45,KMT}
\begin{align} \label{eq:ppxsec}
	\frac{d\sigma_{pp}}{d\Omega_{pp}}=
	\eta^2\frac{d\sigma_{pp}^\textrm{\,t}}{d\Omega_{pp}^\textrm{\,t}},
\end{align}
where
\begin{align}
	\eta=\left(\frac{E_1^\mathrm{t} E_2^\mathrm{t} E_0^\mathrm{t} E_N^\mathrm{t}}{E_1 E_2 E_0 E_N}\right)^{1/2}. \label{eq:mol}
\end{align}

\subsection{Nuclear inputs} \label{sec:input}  
It has been shown in Refs.~\cite{Kra01,Wak17} that a reliable extraction of SF requires the proper uses of bound state wave function constrained by $(e,e'p)$ reaction and nonlocality corrections for both bound and scattering wave functions. In this calculation, the bound state wave function of the struck proton is generated with the well-depth prescription using Woods-Saxon potentials with geometrical parameters taken from $(e,e'p)$ analyses \cite{van88,Kra01}.

For the OPs, we consider in this work three main types: the Schr\"{o}dinger-based microscopic folding and phenomenological potentials, and the Dirac-based phenomenological one. The complex microscopic single-folding potential (MOP) in Brieva-Rook localized form \cite{Min10,Toy13} is obtained with the parameter-free Melbourne $g$ matrix \cite{Amos}, which is constructed based on the Bonn-B free $NN$ interaction \cite{BonnB}. We use the densities calculated with the independent-particle model parametrized by Bohr and Mottelson \cite{BM}. The same folding procedure has been successfully applied in numerous knockout reactions with neutron-rich nuclei in recent years \cite{Tan19,Chen19,Yang21,Oli17,Tang20,Kaw18}. 

For the Schr\"{o}dinger-based phenomenological optical potential, we adopt the well-known global parametrization of Koning and Delaroche (KD) \cite{KD}, which is valid for the energy range from 1 keV to 200 MeV. We note that although the KD potential is originally parametrized for the mass range $24\leq A \leq 209$, it has been shown to provide a good agreement with experimental $p+^{12}$C elastic scattering and reaction cross section data \cite{Au05}. Many studies of $(p,pN)$ reactions in the considered energy region \cite{Cow89,Dev79,Bho76,Sam86} have indicated that the mass and energy systematics of the outgoing OPs affect the TDX significantly more than their detailed shape at a given energy. Combined the above arguments with the small and monotonous mass dependence of the KD potential \cite{KD}, it is completely justified to use the KD OP in this study.      

Finally, we consider the EDAD1 parametrization of the Dirac-phenomenology optical potential in the Schr\"{o}dinger-reduction form \cite{EDAD1}. This complex potential is the mixture of the Lorentz scalar and vector potentials, whose parameters are fitted in the energy range 20--1040 MeV. The SFs extracted from extensive DWIA analyses of $(p,2p)$ reactions with the EDAD1 potential are in good agreement with those from $(e,e'p)$ for a wide range of nuclei \cite{Wak17,Nor20}.        

The nonlocality correction is applied to the bound state wave function and the distorted wave functions generated by the Sch\"{o}dinger-based OPs through the Perey factor \cite{Per63,Tit14}
\begin{align}
	F_\textrm{PR}(R)=\left[1-\dfrac{\mu_{N\mathrm{B}}}{2\hbar^2}\beta^2 U_{p\mathrm{B}}(R)\right]^{-1/2},
\end{align}
where $U_{p\mathrm{B}}$ is the proton-nucleus binding or optical potential and $\beta=0.85$ fm the nonlocality range \cite{Per62}. The Perey correction for the bound state wave function contains a factor to ensure normalization. 

For the distorted wave functions obtained from the Dirac-based OPs, the nonlocality correction is achieved with the Darwin factor \cite{Arn81,Raw85} 
\begin{align}
	F_\textrm{DW}(R)=\left[\dfrac{E_i+U_S(R)-U_V(R)-V_C(R)}{E_i}\right]^{1/2},
\end{align}
where $U_S$, $U_V$, and $V_C$ are the Lorentz scalar, vector, and Coulomb potentials in the Dirac equation, respectively. The physical origin of the Darwin factor and its similar role with the Perey factor were discussed in Ref.~\cite{Raw85}.       
    
The $NN$ scattering cross section is obtained from three different parametrizations of the scattering process. The first one is the $t$ matrix of Franey and Love \cite{FL}, which is based on the SP84 partial-wave analysis \cite{SP84}. Next we consider the cross sections calculated with the Reid93 realistic $NN$ interaction \cite{Reid} based on the Nijmegen partial-wave analysis \cite{Sto93}, which is also used in the CDCC-style transfer-to-the-continuum description of $(p,pN)$ reactions \cite{Mor15}. We note that the same analysis result by the Nijmegen group \cite{Sto93} is also used to construct two other well-known $NN$ interactions, the Argonne v18 \cite{AV18} and CD-Bonn \cite{CDBonn}, which are used in the Faddeev/Alt-Grassberger-Sandhas description of $(p,pN)$ reactions \cite{Cre20,Mec19}. The last $NN$ cross section set considered is directly taken from the SP07 partial-wave analysis \cite{SP07}, which is fitted with $pp$ and $np$ data at energy up to 3 GeV. All calculations in this work use the final energy prescription for the on-shell approximation of the $NN$ cross section. We remark that all the three parametrizations of the $NN$ cross section have been used extensively in modern $(p,pN)$ analyses with various reaction models.           
    
\section{RESULTS AND DISCUSSION}
\subsection{Energy-sharing data at 200 MeV}    

In this part, we investigate the $^{12}$C($p,2p$)$^{11}$B reaction at the incident energy of 200 MeV. The experimental data were measured at the National Accelerator Centre (now the iThemba LABS) in several coplanar energy-sharing configurations \cite{Cow89}, in which the emitted angles are fixed and the outgoing proton energy is varied from 10--170 MeV (see Refs.~\cite{Pil89,Pil-PhD} for more details on the experiment). The data measured in Ref.~\cite{Cow89} are the main focus of this study due to their high-resolution measurement of the outgoing proton energy. The two angle-pair configurations considered are ($\theta_1^L, \theta_2^L$) = ($45^\circ, 35^\circ$) and ($20^\circ, 20^\circ$), with ($\phi_1^L, \phi_2^L$) = ($180^\circ, 0^\circ$). The first configuration lies mostly within a so-called quasifree region with the recoil momentum of the two peaks less than 100 MeV/$c$. The latter one is an example of a nonquasifree kinematic condition with the recoil momenta of the data points mostly around 150 MeV/$c$. At such an extreme kinematic condition, it is expected that the differences between various choices of the nuclear interactions are strongly manifested in the energy-sharing distribution (i.e. the TDX).   

\begin{figure}[bht]\vspace*{0cm}\hspace*{0cm}
	\includegraphics[width=0.85\textwidth]{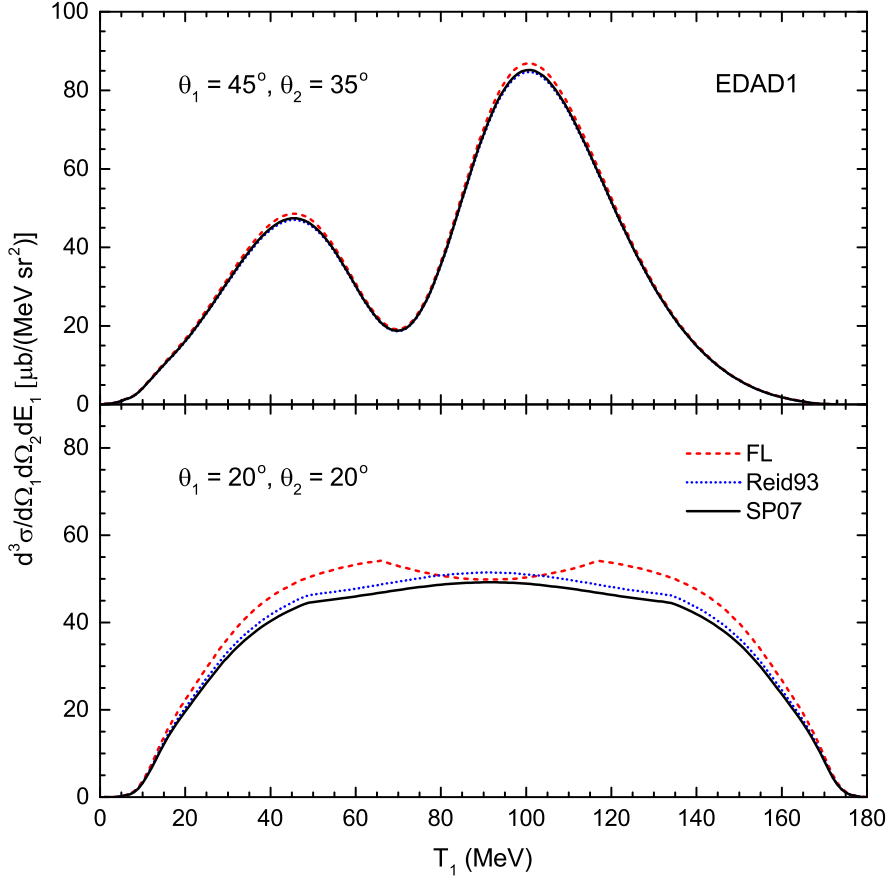}\vspace*{-0.4cm}
	\caption{The TDX of $^{12}$C($p,2p$)$^{11}$B reaction at 200 MeV in the quasifree (upper figure) and nonquasifree (lower figure) angle pairs calculated with the EDAD1 OP and several $NN$ cross sections. These cross sections are the FL (dashed line), Reid93 (dotted line), and SP07 (solid line). All TDXs are calculated with the $p$-state SF = 1.82.} \label{fig:200-NN}
\end{figure}           

Due to the missing mass resolution of 4 MeV in the experimental apparatus \cite{Cow89}, the TDXs reported in this experiment contain both the $3/2^-$ ground and $1/2^-$ 2.13 MeV first excited states of $^{11}$B. For the DWIA analysis of this data, we extracted the $p$-state SF of $^{12}$C by comparing the experimental TDX with the calculated one assuming the transition to the ground state of $^{11}$B. This $p$-state SF is the sum of the proton SFs of $^{12}$C for transitions to two lowest states of $^{11}$B. The use of the single-particle TDX with $^{11}$B in the ground state is justified due to the similarity in shapes of the lowest states TDXs \cite{Wak17,Dev79}. We estimated an uncertainty of 3\% for this extraction procedure of the $p$-state SF. We note that a single SF is used for both scattering geometries, which puts a more severe test on the validity of the DWIA than the analysis in Ref.~\cite{Cow89}.  

\begin{figure}[bht]\vspace*{0cm}\hspace*{0cm}
	\includegraphics[width=0.85\textwidth]{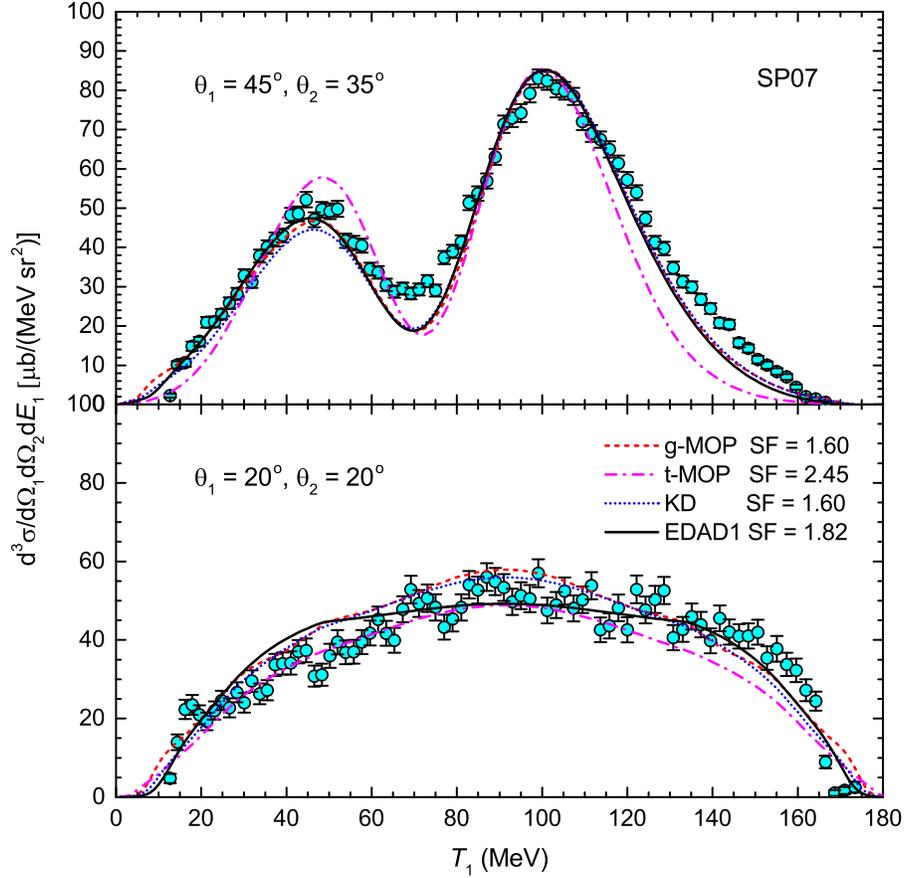}\vspace*{-0.4cm}
	\caption{The TDX of $^{12}$C($p,2p$)$^{11}$B reaction at 200 MeV in angle pairs calculated with the SP07 cross sections. The experimental data \cite{Cow89} are compared with the DWIA calculation using the $g$-matrix (dashed line) and $t$-matrix (dotted-dashed line) folding potentials, and the KD (dotted line) and EDAD1 (solid line) phenomenology OPs. The calculation with each OP use a different $p$-state SF as described in the figure and the text. } \label{fig:200-pot}
\end{figure} 

First, we explore the sensitivity of the TDX on three different empirical $NN$ cross sections in Fig.~ \ref{fig:200-NN}. Due to the small differences in both the shape and magnitude of the calculated TDXs, we did not show the experimental data for the sake of clarity. All three results are calculated with the same EDAD1 OP using the $p$-state SF = 1.82. For the quasifree ($45^\circ, 35^\circ$) case, the differences are less than 3\% at the peak region. At the ($20^\circ, 20^\circ$) geometry, the differences in some region of the TDX is larger than those in the previous case. However, since we use the experimental data of both angle pairs to constrain the SF, it is not affected by the differences in the nonquasifree case. Nevertheless, Fig.~\ref{fig:200-NN} suggests that the TDX at some extreme geometries are strongly sensitive to the subtle feature of the $NN$ interaction. It is noted that besides the final energy prescription, we have tested the initial and average energy prescriptions for the on-shell approximation and found that the magnitudes of the TDXs in the ($20^\circ, 20^\circ$) case are too small to give a realistic SF. Our finding is similar to the one reported in Ref.~\cite{Cow89}. Moreover, the final energy prescription is more physically motivated for $(p,pN)$ reactions since the asymptotic momenta of the two outgoing nucleons are well determined by the measurement. Thus the final energy prescription is used in all calculations in this work.   

In Fig.~\ref{fig:200-pot}, we compared the TDX from DWIA calculation using the $g$-matrix folding ($g$-MOP), KD, and EDAD1 OPs with the experimental data \cite{Cow89}. To explore the medium effect in the TDX, we also investigate the microscopic folding calculation using the Franey-Love $t$ matrix \cite{FL,Amos} ($t$-MOP). All calculations in this figure use the SP07 $NN$ cross section. Our purpose for this figure is to compare the agreements between various OPs and the data in terms of the TDX shape while simultaneously extract the corresponding $p$-state SF through the normalization with the data. The relative differences in terms of magnitude between the TDXs of the considered OPs can be directly inferred from these SFs, which are discussed extensively below. From Fig.~\ref{fig:200-pot}, one can clearly see that the shapes of the energy-sharing distributions calculated with the $g$-MOP, KD, and EDAD1 OPs are almost identical in the ($45^\circ, 35^\circ$) case and are in very good agreement with the measured data. We note that, although not shown here, the more recent version of the Dirac phenomenology OP \cite{Coo09} give similar results to those obtained with the EDAD1 potential. On the other hand, the energy-sharing distribution calculated with the $t$-MOP is much different from the rest and it also does not agree with the ($45^\circ, 35^\circ$) experimental data. This result is not unexpected since the $t$-MOP cannot reproduce the elastic scattering data with the same agreement as the $g$-MOP for incident energies lower than 200 MeV \cite{Are95,Toy13}. Given that the difference between the $g$ matrix and the $t$ matrix is mostly due to the lack of density dependence in the latter one, one can conclude that a proper treatment of medium effect in the distorting OP is critical to reproduce the observed TDX. 

For the ($20^\circ, 20^\circ$) angle pair, there is some deviations in the TDX shape around the $T_1=80$--100 MeV region but the calculated results are in overall good agreement with the data. This agreement is especially significant for the nonquasifree ($20^\circ, 20^\circ$) pair since several key assumptions in the DWIA are expected to be valid only with the quasifree kinematics \cite{Wak17,Kit85}. Within the uncertainty of the data for the ($20^\circ, 20^\circ$) pair, it is not feasible to determine which OP provides the best TDX shape. Nevertheless, these results indicate that, similar to the $NN$ cross section comparison, small differences in the OP for distorted waves can be observed with this kinematic configuration. Such sensitivities of the TDX at this angle pair suggest that it is a suitable experimental configuration to test the validity of various choices of nuclear interaction in the DWIA framework.  

\begin{figure}[bht]\vspace*{0cm}\hspace*{0cm}
	\includegraphics[width=0.77\textwidth]{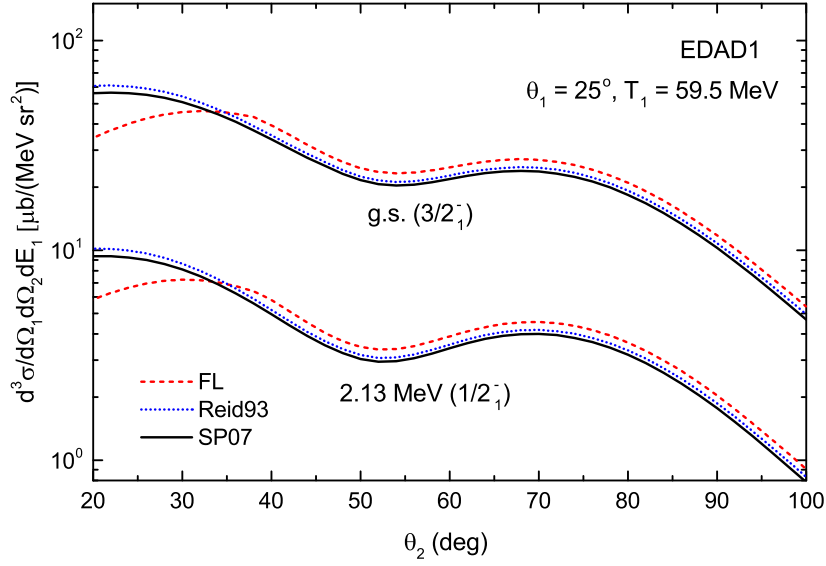}\vspace*{-0.4cm}
	\caption{The TDX of $^{12}$C($p,2p$)$^{11}$B reaction at 100 MeV leading to the ground and first excited states of $^{11}$B calculated with the EDAD1 OP and several $NN$ cross sections. These cross sections are the FL (dashed line), Reid93 (dotted line), and SP07 (solid line). All TDXs are calculated with the SF = 1.00 and 0.15 for the ground and excited state transitions, respectively.} \label{fig:100-NN}
\end{figure}     

While the shape of the TDX can provide information about the angular momentum of the struck nucleon, its SF can be extracted from the magnitude of the TDX. Each calculated TDX in Fig.~\ref{fig:200-pot} corresponding to an OP has been multiplied with a specific $p$-state SF for both angle pairs. The SFs were chosen to provide the best fit between the calculated TDX and the measured one at the $T_1\approx 100$ MeV peak in the ($45^\circ, 35^\circ$) kinematics, where the quasifree condition is fulfilled and the experimental data are measured with good statistics. In this section, we only discuss the best fit SF for the sake of comparison between the OPs. A more careful analysis of the extracted SFs along with their uncertainties is discussed in Sec.~\ref{sec:SF}. 

The extracted $p$-state SF of both the KD and $g$-MOP is 1.60 while the one of the EDAD1 is 1.82. For the considered incident energy of 200 MeV, The EDAD1 SF is in better agreement with the 1.98(11) value from $(e,e'p)$ \cite{Kra01,van88}. Since the TDXs corresponding to these three OPs agree well with the data, the extracted SFs can be regarded as the lower and upper limit of the possible SF with an uncertainty of $6\%$ coming from the choice of OP. As discussed in Ref.~\cite{Wak17}, the difference between the TDX magnitude, or equivalently the SF, calculated by the EDAD1 and those of the Schr\"{o}dinger-based OPs is mainly due to the different impacts of the Darwin and Perey correction factors on the scattering wave functions. Finally, we note that the SF = 2.45 value corresponding to the DWIA calculation with $t$-MOP is unrealistically too large. Therefore, we conclude that the folding potential generated from the free scattering $t$ matrix without any form of medium correction is not advisable to be used in DWIA analysis of proton-induced knockout reaction at the energy below 200 MeV.                
  
\subsection{Angular-correlation data at 100 MeV} 

\begin{figure}[bht]\vspace*{0cm}\hspace*{0cm}
	\includegraphics[width=0.77\textwidth]{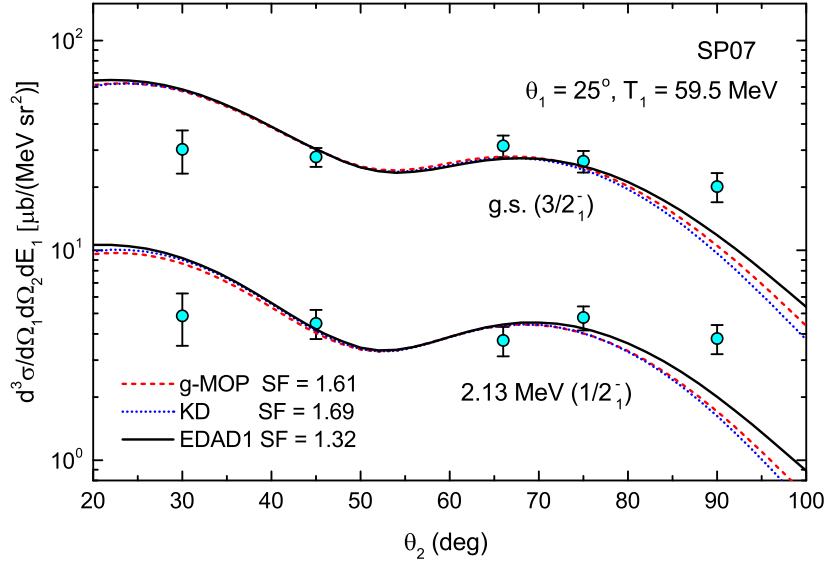}\vspace*{-0.4cm}
	\caption{The TDX of $^{12}$C($p,2p$)$^{11}$B reaction at 100 MeV leading to the ground and first excited states of $^{11}$B calculated with the EDAD1 OP and several $NN$ cross sections. The experimental data \cite{Dev79} are compared with the DWIA calculation using the $g$-MOP (dashed line), KD (dotted line), and EDAD1 (solid line) OPs. The calculation with each OP at each final state use a different SF and the summed $p$-state SFs for each OP are described in the figure and the text. } \label{fig:100-pot}
\end{figure} 

Due to the strong nuclear absorption at certain energies \cite{Wak17,Ber10}, the ideal incident energy range of the $(p,pN)$ reaction is 400--600 MeV based on the energy dependence of the $NN$ total cross section. However, the lower energy limit for the validity of the DWIA formalism is not clearly defined with some suggest it should not be below 200 MeV due to the strong distortion \cite{Kit85}. In this part, we carried out a reanalysis of the 100 MeV $^{12}$C($p,2p$)$^{11}$B data measured at the Indiana University Cyclotron Facility in an asymmetric coplanar configuration \cite{Dev79} to explore the lower energy limit of the current DWIA model. The nuclear inputs used in this calculation as described in Sec.~\ref{sec:input} are considerably more up to date and well constrained than those used in the initial study \cite{Dev79}. The TDX in this measurement is also known as the angular correlation distribution since the scattering angle and kinetic energy of the initial proton after the collision are fixed at ($\theta_1^L, T_1^L$)=($25^\circ, 59.5$ MeV) and the scattering angle of the knocked-out proton is varied from $\theta_2^L=30^\circ$ -- $90^\circ$. Since the transitions to the ground and first excited states of $^{11}$B can be distinguished in these experimental data, the SFs corresponding to these transitions can be separately determined, and the reported $p$-state SF is the sum of these SFs.    

Figure \ref{fig:100-NN} represents the TDXs calculated with the EDAD1 OP and three different empirical $NN$ cross sections. The same SF values of 1.00 and 0.15 are used for the transitions to the ground and excited states of $^{11}$B, respectively. Except for the most forward angles, the shape of these TDXs are very similar to each other. In terms of magnitude, a maximum deviation of 12\% is observed for the TDX calculated with Franey-Love $NN$ cross section from the one of SP07 in most of the angular region. The difference is much larger compared to the one observed at 200 MeV. This reminds us about the high-demanding nature of the kinematic phase space at the low incident energy. 

In Fig.~\ref{fig:100-pot}, we compare the TDX calculated using the $g$-MOP, KD, and EDAD1 OPs with the experimental one \cite{Dev79}. Similar to the analysis at 200 MeV, the SP07 $NN$ cross section is used in the DWIA calculation. For the considered kinematics, the experimental data at $\theta_2^L=45^\circ$ are closest to the recoilless momentum condition. Therefore, the SFs were chosen to provide the best fit with the data at this angle. One can see that for the angular region $\theta_2^L<70^\circ$, the shape of the TDXs calculated with the three OPs are mostly the same. They are also in agreement with the data in the $\theta_2^L=45^\circ$ -- $75^\circ$ range. The further outside this region, the more extreme the kinematics becomes and the calculated result no longer agrees with the data. These deviations from the data can be regarded as the limit of the DWIA formalism at this low incident energy.

The extracted $p$-state SF are 1.61, 1.69, 1.32 for the calculations with the $g$-MOP, KD, and EDAD1 potentials, respectively. This gives an uncertainty of $12\%$ from the choice of OP for this energy. For the considered energy, the SF obtained with the KD OP is closest to the one from $(e,e'p)$. It is interesting to note that in contrast to the $^{12}$C($p,2p$)$^{11}$B reaction at 200 MeV, for this low incident energy the TDX obtained with the EDAD1 potential is larger than those calculated with Schr\"{o}dinger-based OPs, thus giving the smallest SF value. This indicates a stronger energy dependence of the EDAD1 potential compared to those of the KD and $g$-MOP.    

\subsection{Comparison of $p$-state spectroscopic factors}\label{sec:SF}

In this final part of the analysis, we compare the extracted $p$-state SFs in this work from the $^{12}$C($p,2p$)$^{11}$B data at 100 \cite{Dev79} and 200 MeV \cite{Cow89} with those from other reaction and structure studies. Since the $(e,e'p)$ is widely considered to be the best experimental probe for the SF \cite{Aum21}, we use the SF of this method \cite{Kra01,van88} as the reference value to discuss our results. First, we compare our results with those from similar DWIA analyses of the RCNP data at 197 \cite{Nor15,Wak17} and 392 MeV \cite{Nor20}. We note that the DWIA analyses in Refs.~\cite{Wak17,Nor20} of the RCNP data use the same nuclear ingredients as our calculation, namely the EDAD1 OP, SP07 parametrization for $NN$ scattering, and binding potential geometries for the struck proton wave function constrained from the $(e,e'p)$ reaction \cite{Kra01,van88}. The Perey and Darwin nonlocality corrections are also applied for the bound state and scattering wave functions in the same way as our calculation.

\begin{figure}[bht]\vspace*{0cm}\hspace*{0cm}
	\includegraphics[width=0.8\textwidth]{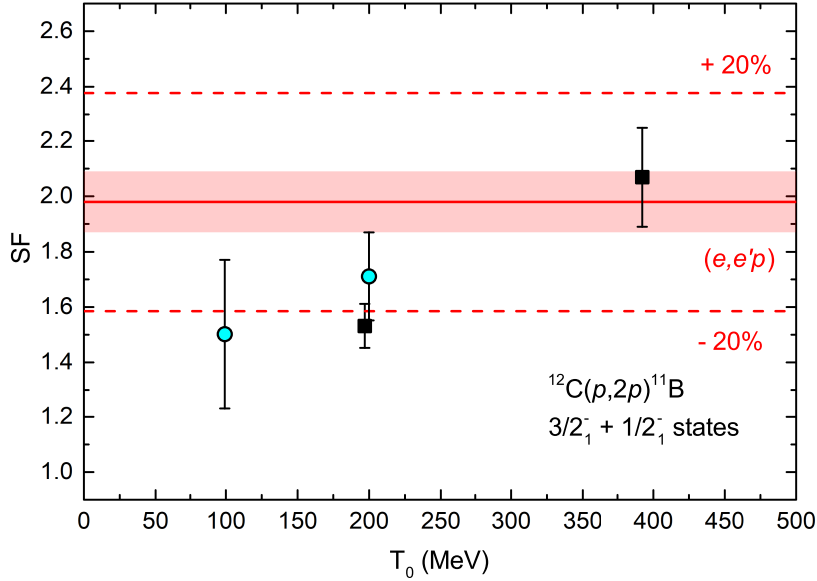}\vspace*{-0.4cm}
	\caption{The $p$-state SFs obtained using $^{12}$C($p,2p$)$^{11}$B data at various incident energies. The SFs from this work at 100 and 200 MeV are shown as dots while those from the analyses at 197 and 392 MeV are represented by squares. The red solid lines and surrounding shaded areas represent the $p$-state SF value and its uncertainty ranges from the $(e,e'p)$ analysis \cite{Kra01,van88}. The $\pm20\%$ deviations from the $(e,e'p)$ results are also shown as dashed lines. The SF values are presented in the text and Table \ref{tab:SF}.} \label{fig:SF}
\end{figure} 

Such similarities in the calculation allow a direct comparison in Fig.~\ref{fig:SF} of the $p$-state SFs extracted from the $^{12}$C($p,2p$)$^{11}$B reaction at four incident energies. In Fig.~\ref{fig:SF} we also show the value from the $(e,e'p)$ reaction along with its uncertainty \cite{Kra01,van88} and $\pm 20\%$ deviation values. The extracted $p$-state SFs with estimated uncertainties in the present works are $1.50(27)$ and $1.71(16)$ at 100 and 200 MeV, respectively. The uncertainties are estimated by taking into account deviations corresponding to different OPs and $NN$ cross section mentioned in the previous section. The $3\%$ uncertainty from the $p$-state normalization procedure for 200 MeV case and the $5\%$ uncertainty coming from the treatment of effective polarization and spin-orbit effect are also included \cite{Cow89}. The total uncertainties as shown in Fig.~\ref{fig:SF} are approximately $18\%$ and $9\%$ for the 100 and 200 MeV energies, respectively. We also list the SFs and their estimated uncertainties in Table \ref{tab:SF}. 

One can see that within the uncertainty range, the SF value at 392 MeV is in complete agreement with the $(e,e'p)$ one while those at lower energies are only consistent with the $20\%$ deviation of the $(e,e'p)$ SF. Figure \ref{fig:SF} also shows that the agreement with the $(e,e'p)$ SF slightly deteriorates with the decrease of energy. This can be explained by the increased influences of several effects, which are not explicitly taken into account in the DWIA model, at energies below 200 MeV. These are the coupled-channel effect \cite{Dev79,Wri78} as well as the off-the-energy-shell \cite{Red70,Wri78} and medium-dependent nature \cite{Wak17} of the $NN$ interaction. These sources of ambiguity are scarcely investigated in the past for the $(p,pN)$ reaction but further studies to tackle them are currently underway. The result in Fig.~\ref{fig:SF} clearly shows that although the $^{12}$C($p,2p$)$^{11}$B reaction analyzed with DWIA framework can provide a reasonable SF down to 100 MeV incident energy, an optimal analysis for this reaction should be done at around 400--600 MeV in a well-chosen quasifree kinematics. From the present analysis and those reported in Ref.~\cite{Wak17}, the 100 MeV energy could be considered as the lowest limit of validity for the partial-wave DWIA formalism \cite{Wak17}. 

\begin{table}[bht]
	\caption{The nucleon SFs of $^{12}$C given by the present study compared with those from other reactions and structure calculations. Details are given in the main text.}
	\begin{tabular}{ccccc}
		\hline
		\multicolumn{5}{c}{$^{12}$C($p,2p$)$^{11}$B } \\		
		Refs.  & Energy (MeV) & g.s. $(3/2^-)$    & 2.13 MeV $(1/2^-)$  & Sum \\
		\hline
		This work & 100 & 1.31(26) & 0.19(3) & 1.50(27) \\
		This work & 200 &       &       & 1.71(16) \\
		\cite{Wak17,Nor15}  & 197 & 1.30(7) & 0.23(3) & 1.53(8) \\
		\cite{Nor20}  & 392 & 1.77(18) & 0.30(3) & 2.07(18) \\
		\hline
		\multicolumn{5}{c}{Other reactions} \\
		Refs.  & Reaction &g.s. $(3/2^-)$    & 2.13 MeV $(1/2^-)$ & Sum \\
		\hline
		\cite{Kra01,van88} & $^{12}$C$(e,e'p)$$^{11}$B   & 1.72(11) & 0.26(2) & 1.98(11) \\
		\cite{Kra01} & $^{12}$C($d,$$^{3}$He)$^{11}$B & 1.72 & 0.27 & 1.99 \\
		\cite{Lee06} & $^{12}$C$(p,d)$$^{11}$C & 2.16(25) &       &  \\
		\cite{Xu19} & $^{12}$C$(p,d)$$^{11}$C & 2.13(25) & 0.48(1) & 2.61(25) \\
		\cite{Li14} & $^{11}$B($^{12}$C$,^{11}$B)$^{12}$C & 2.15(23) &       &  \\
		\hline
		\multicolumn{5}{c}{Structure calculations} \\
		Refs.  & Model & g.s. $(3/2^-)$    & 2.13 MeV $(1/2^-)$ & Sum \\
		\hline
		\cite{Coh67} & \textit{p}-SM  & 2.85 & 0.75 & 3.60 \\
		\cite{Bro02} & \textit{spsdpf}-SM & 3.16 & 0.58 & 3.74 \\
		\cite{Timo13} & STA   & 1.540 & 0.484 & 2.024 \\
		\cite{Cre20} & VMC   & 2.357(12) & 0.868(4) & 3.225(13) \\
		\hline
	\end{tabular}%
	\label{tab:SF}%
\end{table}%

Finally, to provide a broader perspective on the nucleon SF of $^{12}$C, we mention in Table \ref{tab:SF} some SF values from several structure and reaction analyses. For the $(p,2p)$ reaction in the upper part of Table \ref{tab:SF}, in the following discussion we mostly consider the 392 MeV case, which gives the best agreement with the $(e,e'p)$ results. In the middle part of Table \ref{tab:SF}, we present the SFs from recent analyses of transfer reactions. For the nucleon SF of $^{12}$C, the isospin is a good symmetry and thus one can consider the proton and neutron SFs of $^{12}$C are the same. This allows us to include in our comparison the neutron SFs from two systematic $(p,d)$ reactions analyses of Refs.\cite{Lee06,Xu19}. The mean values of the g.s. SFs from transfer reactions \cite{Lee06,Xu19,Li14} are larger than those of $(p,2p)$ at 392 MeV and $(e,e'p)$ by approximately $22\%$. However, we note that within the total uncertainty, the g.s. SF of $(p,2p)$ reaction at 392 MeV are completely consistent with those obtained from other reactions. The seminal work of Kramer \textit{et al.} \cite{Kra01} has shown that by using a consistent treatment of the bound state wave function, nonlocality corrections, and optical potentials, the SFs extracted from transfer and $(e,e'p)$ reactions agree with each other. Therefore, the discrepancies observed in SFs of $^{12}$C between various reactions in Table \ref{tab:SF} are mainly due to the differences in model inputs.    

In the lower part of Table \ref{tab:SF}, we present some SFs from several structure model calculations such as the shell model in $p$ space (\textit{p}-SM) \cite{Coh67} and in much larger \textit{spsdpf} space (\textit{spsdpf}-SM) \cite{Bro02}, the source term approach (STA) \cite{Timo13}, and the \textit{ab initio} Variational Monte Carlo (VMC) model \cite{Cre20}. The $p$-state SFs of $(p,2p)$ and $(e,e'p)$ reaction are about $55\%$ of those from standard shell model calculations \cite{Coh67,Bro02}. This is the well-discussed quenching of SF problem, which is due to the missing of several correlations in the structure calculation (see, e.g. Ref.~\cite{Aum21} for a recent review). It is known that even the VMC calculation does not include all of the correlation required to fully reconcile with the experimentally obtained SF \cite{Cre20,Mec19}. Some contribution from the missing model space can be effectively included in the STA method \cite{Timo13}, which solve an inhomogeneous equation of the overlap function with a suitable effective interaction, thus giving a better agreement with the SFs from $(p,2p)$ \cite{Nor20} and $(e,e'p)$ \cite{Kra01,van88} reactions. We note that the study of SF quenching and its proton-neutron asymmetry dependence is still a very active field of research with many questions yet to be answered \cite{Aum21}.

\section{Summary}
In this work, we have performed a DWIA analysis of $^{12}$C($p,2p$)$^{11}$B data at 100 and 200 MeV. By comparing various well constrained nuclear inputs including the OP and $NN$ scattering cross section, we have estimated the possible ambiguities associated with these components and how they affect the TDX and the extracted SF. Our calculations well reproduce the shape of the TDX for most kinematics at the considered energies. 

A consistent comparison of the extracted $p$-state SFs at several incident energies shows that with carefully chosen kinematics, the SFs obtained with DWIA analyses of $^{12}$C($p,2p$)$^{11}$B reaction with energy down to 100 MeV agree within $20\%$ of deviation from the one given by $(e,e'p)$ reaction. An optimal analysis is suggested to be in the higher energy range of 400--600 MeV with the quasifree kinematic condition. This study further reaffirms the validity of the $(p,pN)$ as a powerful spectroscopic tool. Our results also suggest that some of the keys to improve the $(p,pN)$ analysis at low energy are the OPs specifically optimized for the target and residue nuclei constrained with elastic scattering data in a wide energy region and the $NN$ cross section strictly determined in the considered phase space by the latest data. With the capability of accurately probing the proton and neutron single-particle properties for a wide range of nuclei, including exotic ones, the $(p,pN)$ reaction at intermediate energy is a valuable tool to be used in various existing and upcoming accelerator facilities.   

\section*{Acknowledgments}
We thank the Vietnam MOST for its support through the Physics Development Program Grant No. {\DJ}T{\DJ}LCN.25/18. N.T.T.P also thanks Kazuki Yoshida and Tetsuo Noro for 
valuable discussions about the DWIA calculations.

\bibliographystyle{cip-sty-2019}
\bibliography{refs}

\end{document}